\documentclass[aps,prb,twocolumn,showpacs]{revtex4}
\usepackage{amsmath}

\usepackage{graphicx}

\begin{document}

\title{Transport properties and anisotropy of Rb$_{0.8}$Fe$_2$Se$_2$ single crystals}

\author{Chun-Hong Li$^1$, Bing Shen$^1$, Fei Han$^1$, Xiyu Zhu$^1$, and Hai-Hu Wen$^{1,2}$}

\address{$^1$ National Laboratory for Superconductivity, Institute of Physics and Beijing National Laboratory for Condensed Matter Physics, Chinese Academy of
Sciences, P.O.Box 603, Beijing 100190, China}

\address{$^2$ National Laboratory of Solid State Microstructures and Department of Physics, Nanjing University, Nanjing 210093, China}

\begin{abstract}

Single crystals of Rb$_{0.8}$Fe$_2$Se$_2$ are successfully
synthesized with the superconducting transition temperatures
T$_c^{onset}$ = 31 K and T$_c^{zero}$ = 28 K. A clear anomaly of
resistivity was observed in the normal state at about 150 K, as
found in a similar system K$_x$Fe$_2$Se$_2$. The upper critical
field has been determined with the magnetic field along ab-plane and
c-axis, yielding an anisotropy of about 3.5. The angle dependent
resistivity measured below T$_c$ allow  a perfect scaling feature
based on the anisotropic Ginzburg-Landau theory, leading to a
consistent value of the anisotropy which decreases from about 3.6 at
around T$_c$ to 2.9 at 27 K. Comparing to the anisotropy determined
for Ba$_{0.6}$K$_{0.4}$Fe$_2$As$_2$ and
Ba(Fe$_{0.92}$Co$_{0.08}$)$_2$As$_2$ using the same method, we find
that the present sample is more anisotropic and the Fermi surfaces
with stronger two dimensional characters are expected.

\end{abstract}

\pacs{74.25.fc,  74.70.Xa, 74.62.Bf}
\maketitle

Iron pnictide superconductors have received tremendous attention in
last two years since Kamihara et al. reported superconductivity at
26 K in LaFeAsO$_{1-x}$F$_{x}$. \cite{Kamihara2008} The family of
the FeAs-based superconductors has been expanded rapidly. A typical
example is the (Ba,Sr)Fe$_2$As$_2$ (denoted as FeAs-122) system: the
antiferromagnetic order is suppressed and superconductivity is
induced by either K doping in the Ba or Sr
sites,\cite{2,Canfield,ChuCW2} or Co and Ni doping in the Fe
sites.\cite{Mandrus,Fisher} On the other hand, superconductivity was
also found in the FeAs-based parent phase LiFeAs (denoted as
FeAs-111)\cite{LiFeAs,LiFeAsChu,LiFeAsUK} and Sr$_2$VO$_3$FeAs
(denoted as FeAs-21311).\cite{SrVOFeAs} Compared to these iron
pnictides, FeSe has a more simple structure of only FeSe layers and
no toxic arsenic,\cite{WuMK} which shows superconductivity at 8 K at
ambient pressure and the transition temperature can be increased
dramatically to 37 K under a high pressure.\cite{S. Margadonna}
Moreover, recent report showed that superconducting and magnetic
properties of Fe$_{y}$Se$_{x}$Te$_{1-x}$ not only depend on the
concentration ratio of Se/Te, but also strongly depend on the
interstitial Fe content.\cite{L. Zhang} Additionally, angle resolved
photoemission spectroscopy(ARPES) showed that the normal state of
FeSe$_{0.42}$Te$_{0.58}$ is a strongly correlated metal, which is
significantly different from the FeAs-1111 and FeAs-122
systems.\cite{A. Tamai} Therefore, the FeSe-layered materials
deserve intensive studies for both fundamental physics and potential
applications.

Very recently, superconductivity at around 30 K was reported in
K$_{x}$Fe$_{2}$Se$_{2}$ (denoted as FeSe-122)\cite{chenxiaolong},
where the potassium ions could be intercalated into the Fe$_2$Se$_2$
layers.  This discovery was quickly repeated by other groups with
the nominal composition K$_{0.8}$Fe$_{2}$Se$_{2}$\cite{Yoshikazu
Mizuguchi}. Introducing potassium into the system makes the
structure change from 11-type(P4/nmm) to 122-type(I4/mmm). Up to
now, the system FeSe-122 gives the highest T$_c$ among the
FeSe-layered compounds under ambient pressure. Shortly after that,
Krzton-Maziopa et. al. reported the crystal growth of an analog
compound Cs$_{0.8}$(FeSe$_{0.98}$)$_{2}$.\cite{A Krzton-Maziopa}
Furthermore, Fang et al.\cite{FangMH} synthesized the systematically
doped (Tl,K)Fe$_{2-x}$Se$_2$ and found that the superconductivity
might be in proximity of a Mott insulator. If just counting on the
electron numbers, one would assume that A$_{x}$Fe$_{2}$Se$_{2}$ (A =
alkaline metals) might be a purely electron doped sample. Thus it is
curious to know whether the Fermi surfaces are close to or far
different from their relatives Ba(Sr)Fe$_2$As$_2$. The anisotropy is
one of the important parameters that characterize the electronic
properties. In this work, we report the successful synthesis of a
new compound Rb$_{0.8}$Fe$_2$Se$_2$. The onset and zero-resistivity
transition temperature were estimated to be 31 K and 28 K,
respectively.  We also present the temperature, magnetic field and
angle dependence of resistivity. Our results point to a higher
anisotropy in Rb$_x$Fe$_2$Se$_2$ comparing to electron and hole
doped Ba(Sr)Fe$_2$As$_2$.

\begin{figure}
\includegraphics[scale=0.9]{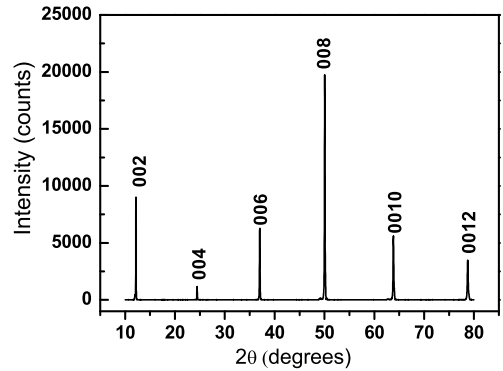}
\caption{ The X-ray diffraction pattern of Rb$_{0.8}$Fe$_2$Se$_2$
crystal indicates that the (00\emph{l}) (\emph{l}=2\emph{n})
reflections dominate the pattern.} \label{fig1}
\end{figure}

Single crystals  were grown from the melt of the mixture of
Rb$_{0.8}$Fe$_2$Se$_2$ using the Bridgeman method. First, FeSe
powders were prepared with high-purity powder of selenium (Alfa,
99.99\%) and iron (Alfa, 99.9\%) by a similar method described in
ref.\cite{11} Then, FeSe and Rb (Alfa, 99.75\%) were mixed in
appropriate stoichiometry and were put into alumina crucibles and
sealed in evacuated silica ampoule. The mixture was heated up to
1030 $^o$C and kept over 3 hours. Afterwards the melt was cooled
down to 730 $^o$C with the cooling rate of 6 $^o$C/h and finally the
furnace was cooled to room temperature with the power shut off. Well
formed black crystal rods were obtained which could be easily
cleaved into plates with flat shiny surfaces. The good $c$-axis
orientation of the crystals has been demonstrated by the X-ray
diffraction (XRD) analysis which show only the sharp (00\emph{l})
peaks. The dc magnetization measurements were done with a
superconducting quantum interference device (Quantum Design, SQUID,
MPMS7). The electrical transport data were collected on the Quantum
Design instrument physical property measurement system (PPMS) with
magnetic fields up to 9 T. The temperature stabilization was better
than 0.1\% and the resolution of the voltmeter was better than 10
nV.

\begin{figure}
\includegraphics[scale=0.95]{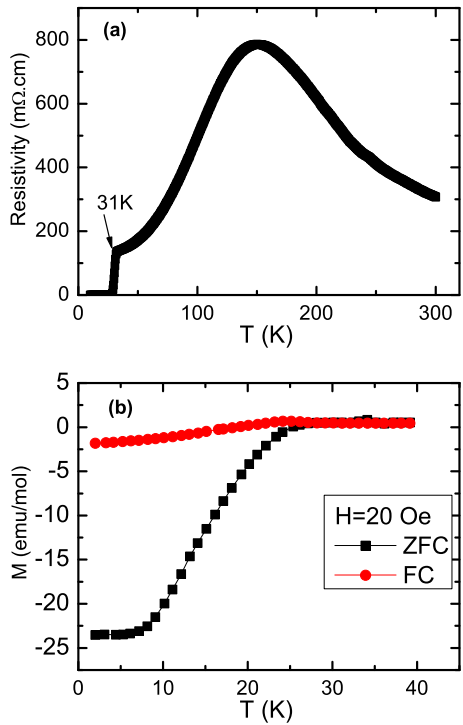}
\caption{(Color online) (a) Temperature dependence of resistivity
for the Rb$_{0.8}$Fe$_2$Se$_2$ crystal at zero field up to 300 K.
A hump of resistivity in the normal state at around 150 K can be
clearly seen. (b) Temperature dependence of dc magnetization for
both zero field cooling (ZFC) and field cooling processes (FC) at
a magnetic field of $H$ = 20 Oe. } \label{fig2}
\end{figure}

Fig.2 (a) shows the temperature dependence of resistivity for a
single crystal of Rb$_{0.8}$Fe$_2$Se$_2$. A superconducting
transition appears at the temperature of 31 K (onset) which is
similar to that of K$_{0.8}$Fe$_{2}$Se$_{2}$.\cite{chenxiaolong} The
bulk superconductivity of our sample is also confirmed by DC
magnetization measurement which is shown in Fig.2 (b), diamagnetism
is clearly observed in both zero-field-cooling and field-cooling
measurement. The relatively broad magnetic transition suggest that
the sample is still inhomogeneous with probably the Rb distributed
non-uniformly. The normal state resistivity of our sample exhibits a
possible semiconductor-to-metal like transition at around 150 K. The
similar behavior was also observed in K$_{0.8}$Fe$_{2}$Se$_{2}$
although in a different temperature region (about 110 K in
K$_{0.8}$Fe$_{2}$Se$_{2}$)\cite{chenxiaolong}. This resistivity
anomaly could also be caused by a structure or magnetic phase
transition which is very typical in the AFe$_2$Se$_2$
superconductors. Further experiments need to be done to clarify the
origin of this transition. It also should be noticed that the
absolute value of normal state resistivity is quite large. The
maximum value exceeds 700 $\;m\Omega\,$cm, which is hundreds of
times larger than that in other typical iron-based superconductors.
This phenomena could be attributed to the semiconductor background
in these iron selenide superconductors.

\begin{figure}
\includegraphics[scale=0.7]{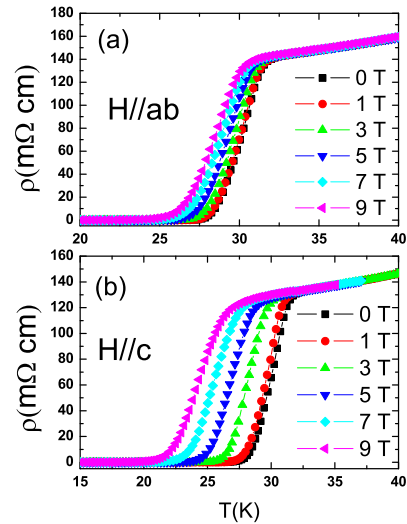}
\caption{(Color online)  The temperature dependence of resistivity
for the Rb$_{0.8}$Fe$_2$Se$_2$ single crystal at zero field and
under magnetic fields of H$//ab$ (a) and H$//c$ (b) up to 9 T with
increments of 2 T. } \label{fig3}
\end{figure}

\begin{figure}
\includegraphics[scale=0.7]{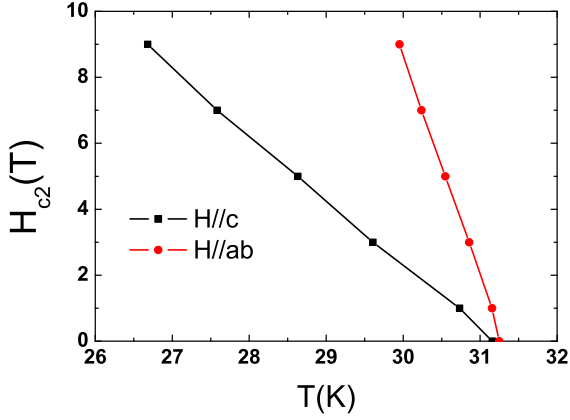}
\caption{(Color online) The upper critical fields of
Rb$_{0.8}$Fe$_2$Se$_2$ single crystal for H$//c$ and H$//ab$
respectively. } \label{fig4}
\end{figure}

The temperature dependence of resistivity from 15 K to 40 K with
different magnetic fields applied along $ab$-plane or $c$-axis are
presented in Fig. 3 (a) and (b). We adopt a criterion of
90\%$\rho_{n}(T)$ to determine the upper critical fields. The
upper critical fields of Rb$_{0.8}$Fe$_2$Se$_2$ are determined in
this way and shown in Figure 4. The upper critical fields H$_{c2}$ exhibit a rather linear temperature dependence for both orientations. Thus we can easily get the
values of the slope for two different directions of applying
fields: $-dH_{c2}^{ab} /dT|_{T_{c}} = $ 6.78T / K, $-dH_{c2}^{c}
/dT|_{T_{c}} = $ 1.98 T / K. The former significantly exceeds the Pauli limit
 1.84 T / K, which may manifest an
unconventional mechanism of superconductivity in this material. With
the Werthamer-Helfand- Hohenberg formula $H_{c2}(0)$ = - 0.69$\times
dH_{c2}/dT|_{T_{c}}T_c$\cite{WHH} and taking $T_{c}$ = 31 K, we can
estimate the values of upper critical fields close to zero
temperature limit:$H_{c2}^{ab}(0)=$ 145 T, $H_{c2}^{c}(0)$ = 42 T.
According to the Lawrence-Doniach model,\cite{LDmodel} the relation
between the anisotropy $\Gamma$ and the upper critical field are
given by

\begin{equation}
\Gamma = (m_{c}/m_{ab})^{1/2} = \xi_{ab}/\xi_{c} =
H_{c2}^{ab}/H_{c2}^{c},
\end{equation}

where H$^{ab} _{c2} $and H$^{c} _{c2} $ are the upper critical
fields with H $\parallel$ab plane and H $\parallel$c axis, m$_c$ and
m$_{ab}$ are the effective masses when the electrons move along
c-axis and ab-plane, $\xi_{ab}$ and $\xi_{c}$ are coherence length
in ab-plane and along c-axis, respectively. From the above data, we
can get the anisotropy of upper critical fields of
Rb$_{0.8}$Fe$_2$Se$_2$, $\Gamma \approx$ 3.5. The anisotropy value
is close to that in K$_{0.8}$Fe$_{2}$Se$_{2}$ ($\Gamma \approx$
3.6).\cite{Yoshikazu Mizuguchi} Compared to the anisotropy in other
FeAs-based superconductors, such as 5 in NdFeAsO$_{1-x}$F$_{x}$,
2-2.5 in Ba$_{1-x}$K$_{x}$Fe$_{2}$As$_{2}$, these values are all
quite small compared to High-$T_c$ cuprates, which indicates an
encouraging application perspective.

Considering the uncertainties in determining the upper critical
field in different formulas and by different criterion, the
anisotropy ratio may subject to a modification. One major concern
was that the zero temperature value $H_{c2}(0)$ was determined by
using the experimental data near $T_c$, this concern can be removed by
the measurements of angular dependent resistivity. According to
the anisotropic Ginzburg-Landau theory, the effective upper
critical field $H_{c2}^{GL}(\theta)$ at an angle $\theta$ is given
by
\begin{equation}
H_{c2}^{GL}(\theta) =
H_{c2}^{ab}/\sqrt{\sin^2(\theta)+\Gamma^2\cos^2(\theta)}.
\end{equation}
The resistivity at different magnetic fields but a fixed
temperature can be scaled with the variable
$H/H_{c2}^{GL}(\theta)$. Thus by adjusting $\Gamma$, the proper
scaling variable
${\widetilde{H}}=H\sqrt{\sin^2(\theta)+\Gamma^2\cos^2(\theta)}$ is
acquired, and then the resistivity measured at different magnetic fields should collapse onto
one curve \cite{Blatter}. Figure 5 presents the data of angular
dependence of resistivity at 27 K, 28 K for the
Rb$_{0.8}$Fe$_2$Se$_2$ single crystal. At each temperature, a
cup-shaped feature centered around $\theta = 90^o$ is observed.
The curves measured at different magnetic fields but at a fixed
temperature are scaled nicely by adjusting $\Gamma$. In this
treatment only one fitting parameter $\Gamma$ is employed in the
scaling for each temperature, so the value of $\Gamma$ is more
reliable than the one determined from the ratio of H$_{c2}^{ab}$
and H$_{c2}^{c}$, which may be affected by using different criterion.
At 27 K and 28 K the anisotropy  $\Gamma$ are found to be 2.9$\pm$0.2 and
3$\pm$0.2, respectively. The results agree very well with the value
determined by the ratio of H$_{c2}^{ab}$ and H$_{c2}^{c}$, which
implies the validity of the values determined in this work. In the
same way, the $\Gamma$ obtained at 29 K and 30 K are
3.3$\pm$0.2 and 3.6$\pm$0.2, which are a little larger than those
at 27 K and 28 K.
\begin{figure}
\center
\includegraphics[scale=0.75]{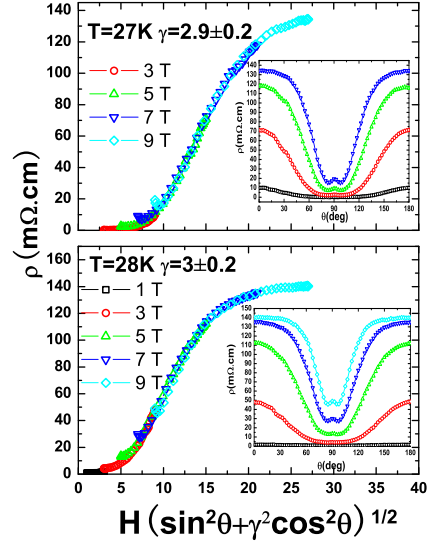}
\caption{(Color online) Scaling of the resistivity versus
$\tilde{H}=H\sqrt{\sin^2(\theta)+\Gamma^2\cos^2(\theta)}$ at 27,
28, 29, 30 K in different magnetic fields. The curves measured at
the same temperature but different magnetic fields are scaled
nicely by adjusting the value of $\Gamma$.  The inset presents the
angular dependence of resistance for the
Rb$_{0.8}$Fe$_{2}$Se$_{2}$ single crystal. }\label{fig5}
\end{figure}

\begin{figure}
\center
\includegraphics[scale=0.75]{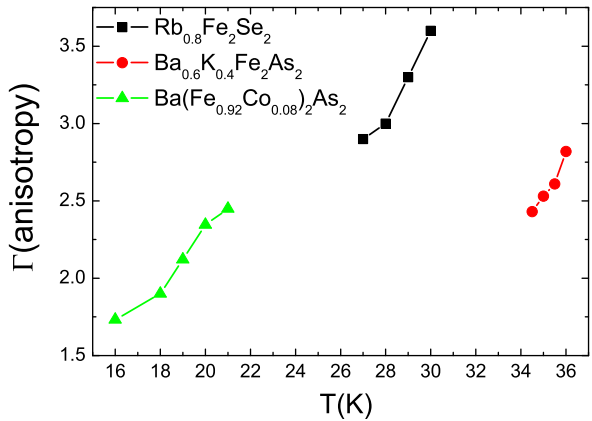}
\caption{(Color online) The comparison of anisotropy  for the
Rb$_{0.8}$Fe$_{2}$Se$_{2}$, Ba$_{0.6}$K$_{0.4}$Fe$_2$As$_2$ and
Ba(Fe$_{0.92}$Co$_{0.08})_2$As$_2$ single crystals.}\label{fig6}
\end{figure}

The anisotropy determined by anisotropic Ginzburg-Landau theory at
different temperatures of Rb$_{0.8}$Fe$_{2}$Se$_{2}$,
Ba$_{0.6}$K$_{0.4}$Fe$_2$As$_2$ and
Ba(Fe$_{0.92}$Co$_{0.08})_2$As$_2$ single crystals are shown in Fig.
6. It is found that, the anisotropy of Rb$_{0.8}$Fe$_{2}$Se$_{2}$
decreases slightly with decreasing temperature. This kind of
temperature dependence of $\Gamma(T)$ is consistent with other
FeAs-122 superconductors, such as Ba$_{0.6}$K$_{0.4}$Fe$_2$As$_2$,
Ba(Fe$_{0.92}$Co$_{0.08})_2$As$_2$, etc.. This may be understood as
the multiband effect, or the effect due to gradual setting in of the
pair breaking by the spin-paramagnetic effect which requires
H$_{c2}^{ab}$ = H$_{c2}^{c}$ in the low temperature and high field
limit. It should be noted that the good scaling behavior suggests a
field-independent anisotropy in the temperature and field range we
investigated. \cite{ZSWang} Compared to MgB$_2$ and cuprate
superconductor, anisotropy of Rb$_{0.8}$Fe$_{2}$Se$_{2}$ is very
small and lower than that of FeAs-1111 family, such as
NdFeAsO$_{1-x}$F$_x$, while it is higher than that in hole and
electron doped FeAs-122 superconductors and similar to that of
KFe$_2$As$_2$ with the same structure.\cite{chengf} It is however
very strange that KFe$_2$As$_2$ and K$_x$Fe$_2$Se$_2$ should reside
in the two terminals of the phase diagram, the former is strongly
hole doped, while the latter is heavily electron doped. The larger
anisotropy in K$_x$Fe$_2$Se$_2$ may suggest a more two dimensional
Fermi surface in this material. So far no angle resolved
photo-emission spectroscopy (ARPES) on the A$_x$Fe$_2$Se$_2$ family
has been reported. The difference between the anisotropy in
Ba$_{0.6}$K$_{0.4}$Fe$_2$As$_2$ and Rb$_{0.8}$Fe$_2$Se$_2$ may hinge
on that the latter has less warped Fermi surface. Our results here
should be stimulating in fulfilling a quantitative calculation and
further studying on the electronic structure of this new family, and
ultimately providing an understanding to the underlying mechanism of
superconductivity.

In conclusion, we successfully fabricate single crystals of
Rb$_{0.8}$Fe$_{2}$Se$_{2}$ with the superconducting transition
temperatures T$_c^{onset}$ = 31 K. A clear anomaly of the
resistivity was observed in the normal state at about 150 K. We also
determined the upper critical fields along ab-plane and c-axis. The
anisotropy of the superconductor determined by the ratio of  H$^{ab}
_{c2} $and H$^{c} _{c2} $ is estimated to be 3.5. The angle
dependent resistivity measured below T$_c$ allow a perfect scaling
based on the anisotropic Ginzburg-Landau theory. The consistent
value of the anisotropy is acquired which decreases from about 3.6
at 30 K around T$_c$ to 2.9 at 27 K. Comparing to the anisotropy
determined for Ba$_{0.6}$K$_{0.4}$Fe$_2$As$_2$ and
Ba(Fe$_{0.92}$Co$_{0.08}$)$_2$As$_2$ using the same method, we
expect that the Fermi surfaces in the new system A$_x$Fe$_2$Se$_2$
is less warped.

This work is supported by the Ministry of Science and Technology of
China (973 project No: 2011CBA00102), the Natural Science Foundation
of China (project number: 51002180), and Chinese Academy of
Sciences.

\end{document}